# Why Bahar and Hausmann Tell Us Nothing About Venezuelan Migration Flows to the United States

Francisco Rodríguez and Giancarlo Bravo[*]

December 17, 2025

## Table of Contents

**Acknowledgments**

*We thank Dorothy Kronick, Carolina Pagliacci, Peter Pedroni, David Rosnick, Leonardo Vera, Mark Weisbrot, and Mark Wohar for valuable comments and Luisa García for excellent research assistance. All errors are our own.*

---

[*] Francisco Rodríguez is a senior research fellow at the Center for Economic Policy and Research and a faculty affiliate at the University of Denver's Josef Korbel School of International Studies. Giancarlo Bravo is the head of research of Oil for Venezuela. Replication data and code for this paper is available at https://doi.org/10.7910/DVN/68B17U.

**Abstract**

Bahar and Hausmann (2025a) claim to find evidence against the hypothesis that oil sanctions on Venezuela lead to increased migration flows to the United States. We show that their findings derive from applying a nonstandard, misspecified Engle-Granger test to first differences. This specification is incorrect because cointegration tests are designed to evaluate relationships between the levels of variables, not their first differences. Since the residuals from regressions of $I(0)$ variables will, under general conditions, be stationary, testing for cointegration between first differences of $I(1)$ variables virtually ensures a spurious finding of cointegration. Using Monte Carlo simulations, we show that the misspecified Bahar-Hausmann test on first differences exhibits a false positive rate of 100 percent. Once the Engle-Granger test is applied correctly to the logarithms of levels, the evidence of cointegration vanishes. The Bahar-Hausmann regressions therefore provide no valid basis for inference about any underlying relationship between migration and Venezuelan oil revenues.

# 1  Introduction

In a working paper and accompanying op-ed published in March of this year, Dany Bahar and Ricardo Hausmann[1] claimed to find evidence of a positive relationship between Venezuelan oil income and migration flows to the United States. Rodríguez, Rosnick, and Bravo[2] showed that these results were driven by a coding error: The authors had incorrectly used a twelfth-difference operator instead of year-on-year seasonal differencing. After correcting this error, the statistical significance of their results disappeared. Further, the observed correlation between Venezuelan migration flows and oil revenues was not robust to the inclusion of controls for labor market tightness in the United States.

Bahar and Hausmann[3] acknowledged that their results lost statistical significance after correcting the coding error yet argued that their positive point estimates still provided evidence against the hypothesis that sanctions — which reduce Venezuelan oil revenues — were a driver of Venezuelan emigration flows to the United States.

---

[1] Bahar and Hausmann (2025a, 2025c).
[2] Rodríguez et al. (2025).
[3] Bahar and Hausmann (2025b).

In this paper, we address a separate problem with the Bahar-Hausmann approach: their incorrect application of cointegration methods.

Bahar and Hausmann's results come from two empirical specifications that are premised on the existence of a cointegrating relationship between migration and oil revenues. Citing results from an Engle-Granger cointegration test, they claim to find evidence of such a relationship. Yet those results come from misapplying a misspecified, nonstandard version of the test and disappear once the error is corrected.

The cointegration results reported by Bahar and Hausmann derive from applying an Engle-Granger test to first differences of their variables of interest. However, the Engle-Granger test is designed to be applied to the *levels* of variables that are integrated of order one, not their first differences. By first differencing the series before applying the Engle-Granger test, Bahar and Hausmann are incorrectly testing for cointegration between two stationary series.[4] Because the test evaluates the stationarity of the residuals from a regression involving potentially cointegrated series, applying it to stationary first differences will virtually guarantee a spurious rejection of the null of no cointegration.

As we show below, once we correct this misspecification, the evidence of cointegration in the Bahar and Hausmann data disappears. This implies that their empirical framework is inadequate and uninformative with respect to the existence of a long-run relationship between oil revenues and migration flows of Venezuelan nationals to the United States.

## 2 Appropriate Specification of Cointegration Tests

### 2.1 General Results

The Dickey-Fuller test statistic for a sample of a time series $Y_t$ is given by the *t*-statistic from regressing $Y_t - Y_{t-1}$ against $Y_{t-1}$ via ordinary least squares (OLS). If $\hat{\rho}_T(Y_t)$ is the OLS estimate of the $AR(1)$ coefficient for a sample size $T$ of $Y_t$ with a resulting estimate of the standard error $\hat{\sigma}_{\hat{\rho}_T}(Y_t)$, then the Dickey-Fuller test statistic can be written as

$$DF_T(Y_t) = \frac{\hat{\rho}_T(Y_t)-1}{\hat{\sigma}_{\hat{\rho}_T}(Y_t)}. \tag{1}$$

---

[4] Vera (2025) was, to the best of our knowledge, the first person to point to this methodological problem in the Bahar-Hausmann paper.

A test statistic close to zero is consistent with the existence of a unit root.

Now let $X_t$ and $Y_t$ denote two $I(1)$ time series, which we observe in the sample $t = 1, \ldots, T$. We say that $X_t$ and $Y_t$ are cointegrated if there exists a vector $\beta = (\beta_1, \beta_2)$ such that the residual

$$\varepsilon_t = \beta_1 X_t + \beta_2 Y_t \tag{2}$$

is stationary. Residual-based tests of cointegration, such as that of Engle and Granger,[5] test the hypothesis that such a vector exists by assessing whether the residual from a regression of $Y_t$ on $X_t$ is stationary.

The Engle-Granger approach consists in constructing a test statistic based on the regression

$$\Delta e_t = (\rho - 1)e_{t-1} + u_t, \tag{3}$$

where $e_t$ are the residuals from an OLS regression of $Y_t$ on $X_t$. The test statistic, which is analogous to (but has a different distribution from) a $t$-statistic for the coefficient estimate of equation (3), is used to test the hypothesis that $\rho = 1$ — that is, that $\varepsilon_t$ has a unit root. Failure to reject the hypothesis that $\varepsilon_t$ has a unit root indicates that the evidence is consistent with the absence of a cointegrating relationship between $Y_t$ and $X_t$. More formally, the Engle-Granger test statistic is

$$EG_T(Y_t, X_t) = DF_T(\varepsilon_T), \tag{4}$$

where $\varepsilon_T$ are the residuals based on the OLS regression of $Y_t$ against $X_t$ for the sample size $T$. A test statistic that is close to zero is consistent with the existence of a unit root in the residuals and therefore is consistent with the absence of cointegration of the underlying variables.

Bahar and Hausmann apply the Engle-Granger test to the first differences of the variables between which they test for cointegration. The Bahar-Hausmann test statistic is thus:

$$BH_T(Y_t, X_t) = EG_{T-1}(\Delta Y_t, \Delta X_t). \tag{5}$$

Note that the sample size is reduced by one as differencing drops the first observation.

---

[5] Engle and Granger (1987).

Rejection of the null of a unit root in this regression implies that the evidence is consistent with the existence of a cointegrating relationship between $X_t$ and $Y_t$, whose residuals are stationary. However, the Engle-Granger test is designed to be applied to the *levels* of $X_t$ and $Y_t$. The reason is that if $X_t$ and $Y_t$ are $I(1)$, then $\Delta X_t$ and $\Delta Y_t$ — the first differences of $X_t$ and $Y_t$ — will be by definition $I(0)$. But then, as we establish below, under general conditions, the residuals from a regression of $\Delta Y_t$ on $\Delta X_t$ will also be stationary. The second step of the Engle-Granger test will therefore be applied to stationary residuals and therefore tend to reject the null of no cointegration, even when no cointegrating relationship exists.

Proposition 1 establishes this result more formally by showing that if we apply the Engle-Granger test for cointegration to the first differences of two non-cointegrated, nonstationary variables, it will reject the null of no cointegration with probability one in sufficiently large samples.

**Proposition 1** Let $X_t$ and $Y_t$ be two non-cointegrated $I(1)$ processes such that $\Delta X_t$ and $\Delta Y_t$ are jointly weakly stationary; ergodic in second moments; and independent with constant means $\mu_x$ and $\mu_y$, constant and positive variances $\sigma_x^2$ and $\sigma_y^2$, and finite moments up to the fourth order. Then, as $T \to \infty$, the probability that the Bahar-Hausmann test falsely rejects the null of no cointegration approaches 1.

**Proof.**

Let $\beta = \frac{\text{cov}(\Delta Y_t, \Delta X_t)}{\text{var}(\Delta X_t)}$ be the linear projection of $\Delta Y_t$ on $\Delta X_t$. Let $\hat{\beta} = \frac{\sum_{t=1}^{T}(\Delta X_t - \overline{\Delta X_T})(\Delta Y_t - \overline{\Delta Y_T})}{\sum_{t=1}^{T}(\Delta X_t - \overline{\Delta X_T})^2}$ and $\hat{\alpha} = \overline{\Delta Y} - \hat{\beta}\overline{\Delta X}$ denote respectively the OLS coefficient and intercept for the regression of $\Delta Y_t$ on $\Delta X_t$ on a sample of size $T$. Ergodicity and finite moments of $\Delta X_t$ and $\Delta Y_t$ imply $\hat{\beta} \xrightarrow{P} \beta$ and $\hat{\alpha} \xrightarrow{P} \alpha$ while independence of $\Delta X_t$ and $\Delta Y_t$ implies $\beta = 0$. Define the residuals of this regression as $e_t = \Delta Y_t - \hat{\alpha} - \hat{\beta}\Delta X_t$ and the corresponding error as $\varepsilon_t = \Delta Y_t - \alpha - \beta\Delta X_t$. Then, $e_t = \varepsilon_t + r_t$, where $r_t = d_\alpha + d_\beta \Delta X_t$, $d_\alpha = \alpha - \hat{\alpha}$ and $d_\beta = \beta - \hat{\beta}$. Note that $E(e_t) = E(\varepsilon_t) + E(r_t) \to 0$ as $E(\varepsilon_t) = 0$ and $|E(rt)| \leq \sqrt{E(d_\alpha^2)} + \sqrt{E(d_\beta^2)}\sqrt{E[(\Delta X_t)^2]} \to 0$ by mean-square consistency of OLS and finite moments of $\Delta(X_t)$. Now, for any fixed lag $k$,

$$Cov(e_t, e_{t-k}) = Cov(\varepsilon_t + r_t, \varepsilon_{t-k} + r_{t-k})$$

$$= Cov(\varepsilon_t, \varepsilon_{t-k}) + Cov(\varepsilon_t, r_{t-k}) + Cov(r_t, \varepsilon_{t-k}) + Cov(r_t, r_{t-k})$$

$$\to Cov(\varepsilon_t, \varepsilon_{t-k}) = Cov(\Delta Y_t, \Delta Y_{t-k}), \tag{6}$$

since

$$|\text{Cov}(\varepsilon_i, r_j)| \leq \sqrt{\text{Var}(\varepsilon_i)\text{Var}(r_j)} \to 0 \tag{7}$$

and

$$|\text{Cov}(r_t, r_{t-k})| \leq \sqrt{\text{Var}(r_t)\text{Var}(r_{t-k})} \to 0 \tag{8}$$

as

$$\text{Var}(r_j) = E(r_j^2) = E(d_\alpha^2) + E(2d_\alpha d_\beta \mu_x) + E(d_\beta^2 \Delta X_j^2) \to 0 \tag{9}$$

and as $d_\alpha \to 0$, $d_\beta \to 0$, and $\Delta X$ has finite second moments. Hence, $e_t$ has a mean that converges to zero and an autocovariance function that converges to a limit independent of $t$, making $e_t$ asymptotically weakly stationary. Let $\tilde{e}_t$ be a zero-mean weakly stationary process that is ergodic in second moments with $Cov(\tilde{e}_t, \tilde{e}_{t-k}) = Cov(\Delta Y_t, \Delta Y_{t-k})$, and let $\gamma_1 = \frac{Cov(\Delta \tilde{e}_t, \tilde{e}_{t-1})}{\text{Var}(\tilde{e}_{t-1})}$ be the linear projection coefficient of $\Delta \tilde{e}_t$ on $\tilde{e}_{t-1}$. Let $\hat{\gamma}_1$ be the OLS coefficient estimate of the regression

$$\Delta e_t = \gamma_1 e_{t-1} + u_t, \tag{10}$$

with $u_t$ being a regression disturbance term. Let $t^s$ denote the associated $t$-statistic, which is also the Bahar-Hausmann test statistic. Then

$$BH_T = t^s = \frac{\hat{\gamma}_1}{\hat{\sigma}_{\hat{\gamma}_1}}, \tag{11}$$

where

$$\hat{\sigma}_{\hat{\gamma}_1} = \left( \frac{\sum_{t=1}^{T}(\Delta e_t - \hat{\gamma}_1 e_{t-1})^2}{(T-1) \sum_{t=1}^{T} e_{t-1}^2} \right)^{\frac{1}{2}}. \tag{12}$$

Since $\tilde{e}_t$ is weakly stationary, $\gamma_1 = \frac{Cov(\Delta \tilde{e}_t, \tilde{e}_{t-1})}{\text{Var}(\tilde{e}_t)} - 1 = \rho(1) - 1$ — where $\rho(1)$ denotes the lag-1 autocorrelation — is independent of $t$. Note that if $\rho(1) = 1$, then the Cauchy-Schwarz inequality would hold as an equality, implying $\tilde{e}_t = \tilde{e}_{t-1}$ and $\text{Var}(\Delta \tilde{e}_t) = 0$. However, $\text{Var}(\Delta \tilde{e}_t) = \text{Var}(\Delta^2 Y_t) > 0$, implying that $\rho(1) < 1$ and $\gamma_1 < 0$. Since $\tilde{e}_t$ is weakly stationary and ergodic in second moments and $e_t$ is asymptotically equivalent to $\tilde{e}_t$, then the sample cross-moments of $e_t$ converge to those of $\tilde{e}_t$, OLS is consistent, $\hat{\gamma}_1 \to \gamma_1 < 0$, $\left( \frac{\sum_{t=1}^{T}(\Delta e_t - \hat{\gamma}_1 e_{t-1})^2}{T-1} \right)^{\frac{1}{2}} \to \sigma > 0$, and given that $\sum_{t=1}^{T} e_{t-1}^2 = O(T)$, $\hat{\sigma}_{\hat{\gamma}_1} \to 0$.

Therefore, $BH_T \to -\infty$. However, we know that the critical value for the Engle-Granger test statistic $Z_t$ converges to a finite value.[6] Therefore, $P(BH_T < Z_t) \to 1$.

Proposition 1 establishes that the probability of rejecting the null of no cointegration using the Bahar-Hausmann approach — that is, (incorrectly) applying the Engle-Granger test to first differences — approaches one even when the series are not cointegrated.

## 2.2  Monte Carlo Simulations

The previous section dealt with the large sample properties of the Bahar-Hausmann approach. This leaves open the question of what the properties of their test are in small samples, such as the one that they used where $N = 48$.

To assess the small-sample properties of their approach, we carried out Monte Carlo simulations under a simple specification in which $x_t$ and $y_t$ are both independent unit root processes that are not cointegrated. Specifically,

$$x_t = x_{t-1} + u_{1t} \quad \text{and} \quad y_t = y_{t-1} + u_{2t},$$

where $u_{1t}, u_{2t} \sim \mathcal{N}(0,1)$ and $\text{Cov}(u_{1t}, u_{2t}) = 0$.

We then applied the Engle-Granger test both in levels and in first differences over 1,000 replications of these series with $N = 48$. The results are summarized in **Table 1** and **Figure 1**. As expected, when the test is run on levels, it rejects the null of no cointegration using MacKinnon's 5 percent critical value in 5.3 percent of simulations and produces an average $t$-statistic of −2.078. This illustrates the well-known result that $t$-statistics applied to equation (2) tend to lie in absolute value above critical values from a standard $t$-distribution but below the MacKinnon critical values used in the Engle-Granger procedure.

However, the second column of Table 1 shows that if one uses the misspecified Bahar-Hausmann approach of running the Engle-Granger test on first differences, the average $t$-statistic drops to −6.956, with an absolute value well above that of the Engle-Granger MacKinnon 5 percent critical value of −3.469. As a consequence, the Bahar-Hausmann test rejects the null of no cointegration — despite the series not being cointegrated — in all of the 1,000 replications, implying a false positive rate of 100 percent.

---

[6] Phillips and Ouliaris (1990, Theorem 4.2); MacKinnon (2010, Table 1).

Table 1: Monte Carlo Simulation Results Under No Cointegration

|  | Levels (undifferenced) | First differences |
| --- | --- | --- |
| Test statistic | −2.078 | −6.956 |
|  | (0.843) | (1.055) |
| Rejection rate | 5.3% | 100.0% |

*Notes:* Table shows the average test statistics and standard errors (in parentheses) from 1,000 simulations, with the Engle-Granger two-stage test carried out on two variables, both of which are modeled as independent unit root processes that are not cointegrated.

Figure 1: Distribution of Engle-Granger Test Statistics Using Level Versus First-Differenced Data

*Monte Carlo Simulations Under No Cointegration*

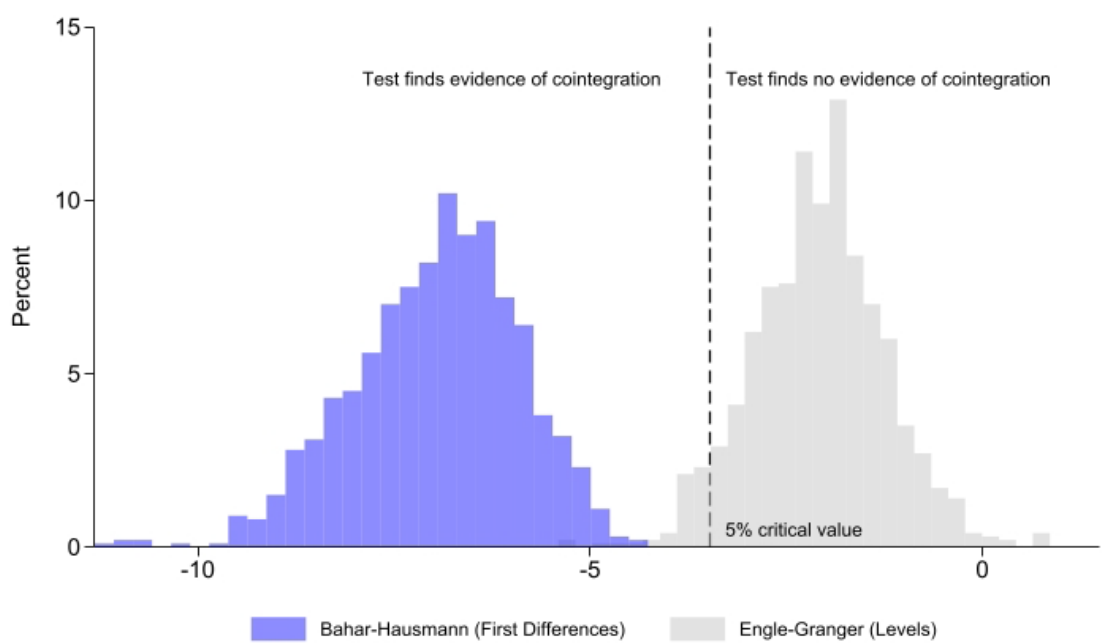

*Note:* Series are constructed to be non-cointegrated random walks with white noise residuals. "Test finds (no) evidence of cointegration" refers to Engle-Granger test (not) rejecting hypothesis of no cointegration.

## 3    Empirical Results

Bahar and Hausmann apply an Engle-Granger cointegration test to the first differences of encounters of Venezuelan nationals at the southwest US border and Venezuelan oil income (estimated as the product of the price times the country's production) and report a test statistic of −7.923, with an absolute value well above that of the 1 percent MacKinnon critical value of −4.145.[7]

Based on their claim of having found cointegration, they go on to estimate two specifications. The first one is a logarithmic specification with fixed month and year effects:

$$\log crossings_{my} = \beta_1 \log oil_{my} + \eta_m + \gamma_y + \varepsilon_{my}, \qquad (13)$$

where *crossings* refers to apprehensions of Venezuelan nationals at the US southern border, *oil* refers to the measure of oil revenues, and *m* and *y* are, respectively, month and year fixed effects, while the subindices *m* and *y* refer to the month and year of the data. The second specification is an autoregressive distributed lag (ARDL) given by

$$\Delta_y \log crossings_{my} = \beta_0 + \beta_y \Delta_y \log oil_{my} + \beta_1 \Delta_y \log crossings_{my-1} +$$
$$\beta_2 \Delta_y \log oil_{my-1} + \beta_3 ECT_{my-1} + \varepsilon_{my}, \qquad (14)$$

where $\Delta_y$ denotes the year-over-year difference of a variable,

$$\Delta_y x_t = x_t - x_{t-12}; \qquad (15)$$

that is, the difference between its level at time *t* and its level in the same month of the previous year, $t - 12$.

There are thus two sources of misspecification in the Bahar-Hausmann application of the Engle-Granger test. The first one, which we already discussed in the previous section, has to do with inappropriately applying the test to first differences instead of levels. The second one has to do with the inconsistency of testing for cointegration on the untransformed variables and then estimating empirical specifications premised on cointegration in the logarithms of those variables. The reason for this is that cointegration among a set of variables does not guarantee, or is not implied by, cointegration in the logarithmic transforms of that set of variables.

This contradiction is particularly relevant in the case of equation (14). This equation controls for an error correction term, $ECT_{my-1}$, which is the series of lagged residuals from an OLS regression in logarithms. But Bahar and Hausmann have not claimed to have found cointegration in logarithms; they claim to have found cointegration in the untransformed series. The validity of $ECT_{my-1}$ as a regressor depends on the existence of a cointegrating regression in logarithms, not in the untransformed values.

---

[7] Bahar and Hausmann actually report the test statistic as −7.903. Yet this is an apparent typographical error, as their own replication code verifies that the test statistic is −7.923.

Equation (13) is also problematic because it reveals an internal inconsistency in Bahar and Hausmann's estimation strategy. If there is a cointegrating relationship between log(*crossings*) and log(*oil*), then that relationship is superconsistently estimated through OLS.[8] But then there is little reason to estimate equation (13), as the long-run relationship between the variables would be given by the OLS estimate obtained in the first stage of the Engle-Granger test or could be obtained through more efficient estimation methods, as discussed below. Alternatively, if the reason to estimate equation (13) is the belief that the cointegrating relationship is affected by seasonality or structural breaks, then these would need to be accounted for in the cointegration tests.[9]

Further, even if we accept interpreting equation (13) as a cointegrating relationship, we know that the standard errors of such a relationship are inconsistently estimated via OLS, which invalidates any hypotheses tests carried out by Bahar and Hausmann using the OLS standard errors. The commonly accepted practice to carry out statistical inference in these cases is to use other estimation methods — such as dynamic OLS or fully modified OLS — capable of correcting these biases.[10]

Remarkably, the OLS estimate of the cointegrating vector as implemented by Bahar and Hausmann — that is, in first differences of the untransformed series — yields a negative coefficient, contrary to their central claim that oil revenues are not negatively related to migration flows (see **Appendix Table A.1**).**Table 2** shows the results of applying the Engle-Granger cointegration test to the undifferenced logarithms of border encounters and oil income measures. The table reports results for the three different measures of oil revenues used by Bahar and Hausmann (oil income, oil prices, and oil production). It also presents results for three different variants of the Engle-Granger test: the basic test, an augmented Engle-Granger test controlling for 12 monthly lags, and an augmented Engle-Granger test controlling for 12 monthly lags and a trend. None of the nine test statistics associated with these specifications reject the null of no cointegration.

---

[8] Hamilton (1994, p. 587).

[9] For there to be a long-run cointegrating relationship among $y_t$, $x_t$, and a set of seasonal dummy variables $D_{m,t}$, where $m = 1, \in 1, M$, there must exist constants $\alpha$, $\beta$, and $\delta_m$ such that the residual $\varepsilon_t = y_t - \alpha - \beta x_t - \sum_{m=1}^{M} \delta_m D_{m,t}$ is stationary. But for there to be a cointegrating relationship between $y_t$ and $x_t$ alone, it must be the case that $n_t = y_t - \alpha_1 - \beta_1 x_t = \varepsilon_t + (\alpha - \alpha_1) + (\beta - \beta_1)x_t + \sum_{m=1}^{M} \delta_m D_{m,t}$ is stationary. Since $x_t$ is $I(1)$, this requires $\beta = \beta_1$ and $\delta_m = 0 \ \forall m$, negating the assumption of seasonality. It follows that if there are nonzero seasonal effects, there is no cointegrating relationship between $y_t$ and $x_t$ alone.

[10] Phillips and Hansen (1990); Stock and Watson (1993).

Table 2: Engle-Granger Tests for Cointegration, Logarithmic Transform

|  |  | Log income | Log price | Log production |
|---|---|---|---|---|
| Engle-Granger | Test statistic | −2.72 | −2.40 | −2.71 |
|  | Number of observations | 47 | 47 | 47 |
|  | Bartlett white noise statistic | 1.18 | 1.35* | 0.75 |
| Aug. Engle-Granger (12 lags) | Test statistic | −1.83 | −0.92 | −1.46 |
|  | Number of observations | 35 | 35 | 35 |
|  | Bartlett white noise statistic | 0.41 | 0.58 | 0.39 |
| Aug. Engle-Granger (12 lags + trend) | Test statistic | −2.19 | −2.32 | −1.63 |
|  | Number of observations | 35 | 35 | 35 |
|  | Bartlett white noise statistic | 0.37 | 0.36 | 0.40 |

*Notes:* The table shows Engle-Granger test statistics under alternative specifications and choice of dependent variables. All series are in logarithms. Bartlett white noise statistics are included for residual autocorrelation diagnostics. Asterisks denote significance at *** 1%, ** 5%, * 10%

We present several additional robustness tests in **Appendix Tables A.2–A.4**. In particular, we report results for Engle-Granger and augmented Engle-Granger tests using the untransformed values of the variables. Most of these specifications also fail to find evidence of cointegration. We also include diagnostic Dickey-Fuller and augmented Dickey-Fuller tests that are commonly run prior to implementing cointegration tests. While it is standard practice to use these tests to verify whether variables are nonstationary before applying cointegration tests, Bahar and Hausmann do not present any results of nonstationarity tests. For both the log and level specifications, we find inconclusive evidence of nonstationarity.[11]

In sum, the cointegration test that corresponds to Bahar and Hausmann's empirical specification should have been conducted in undifferenced logarithms, not in the first differences of the untransformed series that they used. Once one applies the test that is consistent with their empirical specification, evidence of cointegration disappears. This renders invalid both of their estimation exercises and suggests that any results obtained by them could be the product of misspecification, leading to spurious associations between time series subject to stochastic trends.

---

[11] However, see Reed and Smith (2017), who note that standard unit-root tests may over-reject the null of a unit root in the presence of cointegration.

## 4 Conclusions

This research paper has argued that the claim made by Bahar and Hausmann[12] — that sanctions on Venezuela's oil industry are not a driver of migration flows to the United States — finds no support in their own data. We have shown that Bahar and Hausmann's results were based on their use of a nonstandard and misspecified test to evaluate the existence of a long-run cointegrating relationship between Venezuelan migration to the United States and Venezuelan oil revenues. Once their mistake is corrected, evidence of cointegration disappears, rendering their empirical approach invalid.

We have established that when the Bahar-Hausmann test is applied to independent weakly stationary series, the probability that it will find evidence of cointegration 100 percent of the time, even when such a relationship does not exist, approaches one as the sample size increases. We also provide Monte Carlo simulations on time series of the length used by Bahar and Hausmann that yield a false positive rate of 100 percent for two independent non-cointegrated unit root processes. These results imply that the Bahar-Hausmann approach is guaranteed to find evidence of cointegration even when no such relationship exists. We also showed that when one runs cointegration tests on the levels of logarithms — the only formulation consistent with the empirical specifications adopted in their time series regressions — there is no evidence of cointegration in these tests. This renders their empirical approach, including both the level regressions and ARDL specifications, as inherently misspecified and therefore uninformative about the true nature of the relationship.

These results are not surprising. As discussed in greater detail in Rodríguez, Rosnick, and Bravo,[13] there are reasons to be skeptical about the use of data on US border encounters of Venezuelan nationals to assess hypotheses about the effects of sanctions on migration flows. A large share of Venezuelans who attempt to enter the United States today have spent several years living outside of Venezuela, often in countries like Colombia, Peru, or Mexico. It is unclear why their decisions to migrate to the United States would be affected by fluctuations in the resources under the control of Nicolás Maduro's government. As we also show there, any residual correlation between migration and oil prices is explained by the fact that oil prices were strongly correlated with economic conditions in the United States in a period that includes the COVID pandemic. Strikingly, Bahar[14] argues that US labor market conditions are a key determinant of border crossings, yet Bahar and Hausmann make no effort to control for this effect.

---

[12] Bahar and Hausmann (2025a).
[13] Rodríguez et al. (2025).

Moreover, directly testing for the impact of oil revenues on migration flows fails to capture the broader economic dynamics likely to influence migration decisions. Since 2012, Venezuela has suffered the largest documented economic collapse to occur outside of war, and the dramatic increase in emigration during this period suggests that deteriorating economic conditions have been a central factor. In this context, it appears more reasonable to first assess the impact of economic sanctions on Venezuelan economic performance and to separately assess how deteriorating living standards have affected migration decisions.[15]

Bahar and Hausmann have proposed an interesting hypothesis: that expectations of regime change may themselves be a major driver of migration. This idea is not necessarily inconsistent with the view that deteriorating incomes — partly due to sanctions — have contributed to rising emigration. The possibility that migration decisions are shaped by potential migrants' expectations about future economic and political conditions appears inherently plausible.

More empirical work is certainly needed to better understand the extent to which economic and noneconomic factors have contributed to Venezuelans' emigration decisions and to clarify how past and current conditions — as well as expectations of change — shape these choices. While we believe that Bahar and Hausmann's empirical work has failed to yield informative answers to these questions, we consider the questions they have raised to be highly relevant and deserving of further empirical investigation.

---

[14] Bahar (2025).
[15] See Rodríguez (2024) for an example of this approach.

# 5 Appendix Tables

Table A.1: Bahar-Hausmann Implementation of Engle-Granger Test

First Stage: Dependent Variable Is Δ Encounters

|  | Coefficient | Standard error | P-value |
|---|---|---|---|
| ΔOil income | −1.790 | 10.901 | 0.87 |

Second Stage: Dependent Variable Is Δ Residuals

|  | Test statistic | 1% critical value |
|---|---|---|
| $Z(t)$ | −7.923 | −4.145 |

*Notes:* Two-stage Engle-Granger test applied to first-differenced variables. The second-stage test statistic corresponds to the augmented Dickey-Fuller test on the residuals of the first-stage regression.

Table A.2: Engle-Granger Tests for Cointegration, Untransformed Series

|  |  | Income | Price | Production |
|---|---|---|---|---|
| Engle-Granger | Test statistic | −3.54** | −3.34* | −3.69** |
|  | Number of observations | 47 | 47 | 47 |
|  | Bartlett white noise statistic | 0.76 | 0.75* | 0.73 |
| Aug. Engle-Granger (12 lags) | Test statistic | −1.59 | −1.40 | −1.51 |
|  | Number of observations | 35 | 35 | 35 |
|  | Bartlett white noise statistic | 0.35 | 0.33 | 0.25 |
| Aug. Engle-Granger (12 lags + trend) | Test statistic | −1.90 | −1.95 | −1.90 |
|  | Number of observations | 35 | 35 | 35 |
|  | Bartlett white noise statistic | 0.28 | 0.30 | 0.25 |

*Notes:* The table shows Engle-Granger test statistics under alternative specifications and choice of dependent variables. All series are untransformed. Bartlett white noise statistics are included for residual autocorrelation diagnostics. Asterisks denote significance at *** 1%, ** 5%, * 10%.

Table A.3: Dickey-Fuller and Augmented Dickey-Fuller Tests, Logarithmic Transform

|  |  | Log encounters | Log income | Log price | Log production |
|---|---|---|---|---|---|
| Dickey-Fuller | Test statistic | −3.14** | −4.36*** | −3.51** | −2.85* |
|  | Number of observations | 47 | 47 | 47 | 47 |
| ADF (12 lags) | Test statistic | −1.72 | −2.80* | −1.86 | −0.85 |
|  | Number of observations | 35 | 35 | 35 | 35 |
| ADF with trend (12 lags) | Test statistic | −2.12 | −2.73 | −2.06 | −3.86** |
|  | Number of observations | 35 | 35 | 35 | 35 |

*Notes:* Test statistics are based on the Dickey-Fuller and augmented Dickey-Fuller (ADF) tests applied to logarithmic series. Asterisks denote significance at *** 1%, ** 5%, * 10%.

Table A.4: Dickey-Fuller and Augmented Dickey-Fuller Tests, Untransformed Series

|  |  | Encounters | Income | Price | Production |
|---|---|---|---|---|---|
| Dickey-Fuller | Test statistic | −3.26** | −2.50 | −2.62* | −1.93 |
|  | Number of observations | 47 | 47 | 47 | 47 |
| ADF (12 lags) | Test statistic | −1.36 | −2.83* | −1.89 | −0.69 |
|  | Number of observations | 35 | 35 | 35 | 35 |
| ADF with trend (12 lags) | Test statistic | −2.23 | −2.60 | −1.97 | −3.14 |
|  | Number of observations | 35 | 35 | 35 | 35 |

*Notes:* Test statistics correspond to Dickey-Fuller and augmented Dickey-Fuller (ADF) tests on level (nonlogged) data. Asterisks indicate significance at *** 1%, ** 5%, * 10%.